\documentclass[aps,twocolumn]{revtex4}

\usepackage{epsfig}
\usepackage{color}

\begin{document}

\title{Stochastic thermodynamics and entropy production \\
of chemical reaction systems}
\author{T\^ania Tom\'e and M\'ario J. de Oliveira}
\affiliation{Instituto de F\'{\i}sica,
Universidade de S\~{a}o Paulo\\
Rua do Mat\~ao, 1371,
05508-090 S\~{a}o Paulo, SP, Brazil}
\date{\today}

\begin{abstract}

We investigate the nonequilibrium stationary states of systems consisting
of chemical reactions among molecules of several chemical species.
To this end we introduce and develop
a stochastic formulation of
nonequilibrium thermodynamics of chemical reaction systems based on
a master equation defined on the
space of microscopic chemical states, and on appropriate definitions of
entropy and entropy production, 
The system is in contact with a heat reservoir, and is
placed out of equilibrium by the contact with particle reservoirs.
In our approach, the fluxes of various types, such as
the heat and particle fluxes, play a fundamental role in characterizing
the nonequilibrium chemical state. 
We show that the rate of entropy production in the stationary 
nonequilibrium state is a bilinear form in
the affinities and the fluxes of reaction, 
which are expressed in terms of rate constants and transition rates,
respectively.
We also show how the description in terms of microscopic states
can be reduced to a description in terms of the numbers of
particles of each species, from which follows the chemical master equation.
As an example, we calculate the rate of entropy production
of the first and second Schl\"ogl reaction models. 

\end{abstract}

\maketitle

\section{Introduction}

Chemical reaction systems are understood as systems in which one or more
chemical reactions take place \cite{avery1974,yeremin1979,houston2001}. 
A chemical reaction can be as simple as a unimolecular reaction 
in which one molecule of a chemical species transforms into another
molecule of a distinct chemical species,
or it can be a complex process in which molecules of distinct species
dissociate and recombine producing molecules of other species.
In the process, a certain amount of energy is absorbed or released. 
A relevant feature of reactions taking place in a vessel is their intrinsic
stochastic nature which give rise to random fluctuations on the quantities
that describe the reactive system. An appropriate approach that
takes into account this feature is the description of the time evolution
of reactive systems by a continuous time Markov process 
\cite{nicolis1977,kampen1981,tomebook2015}.
That is, one assumes that the time evolution of the system is governed
by a master equation and that the reaction rates are identified as the
transition rates defining the stochastic process. 

Usually the stochastic description of reactive systems is made in terms
of the numbers of particles of each species, identified as stochastic
variables, and the stochastic process is 
understood as a stochastic trajectory in the space spanned by the
numbers of particles of each species. The process is 
a birth and death process in several variables
and the corresponding master equation is called chemical master equation
\cite{nicolis1977,kampen1981,tomebook2015,mcquarrie1967,kurtz1972,
gillespie1992,jiuli1984,mou1986,gillespie2000,horowitz2015,ge2016}. 
Here we consider a more general stochastic approach in which the
state of the system consists of the set of microscopic states such as
that used in stochastic lattice models \cite{tomebook2015}.
The idea of using microscopic states
\cite{qian2016a}
is not new but here we consider a more complete and explicit treatment
by the use of a microscopic description in terms
of microscopic chemical states, 
defined by the set of variables that denote the chemical
species of {\it each} molecule.
Under some circumstances, as we shall see, it is possible to pass
from the microscopic description to the description in
terms of the number of particles.

Our general stochastic approach
is fully connect to thermodynamics and in this sense it 
can be understood as a stochastic thermodynamics of reactive systems.
Stochastic thermodynamics
\cite{jiuli1984,mou1986,horowitz2015,ge2016,qian2016a,tome2006,tome2010,
tome2012,tome2015,zia2006,schmiedl2007,zia2007,
seifert2008,esposito2009,broeck2010,zhang2012,ge2012,
esposito2012,seifert2012,luposchainsky2013,wu2014}
assumes that the time evolution of
a system is a Markovian process, such as that described by a master
equation or by a Fokker-Planck equation, and is based on two assumptions
concerning entropy. The first is that the entropy has the Gibbs form
and the second is that the production of entropy is related to
the probabilities of the direct and reverse trajectories.
This definition of entropy production is explicitly
tied to the dynamics and, at first sight, seems to have
no reference to the energetics, 
in contrast to thermodynamics \cite{horowitz2015}.
However, from the definition of entropy production
it is in fact possible, as we will show
below, to connect the heat flux, the flux of energy plus
the flux of work, to the flux of entropy, providing the 
consistency of stochastic thermodynamics
\cite{tome2006,tome2010,tome2012,tome2015}.

Here we will be concerned mainly with the steady state,
which might be an equilibrium or a nonequilibrium state.
In the latter case each reaction taking place in a vessel
will be in general either shifted
to the products or to the reactants, a situation in which
entropy is continuously being generated and fluxes of several types
such as the flux of particles and the flux of entropy are occurring.
In the case of equilibrium all fluxes vanish, including the entropy
flux, a result that is the hallmark of what is meant by thermodynamic
equilibrium. The vanishing of fluxes is a direct consequence of the
microscopic reversibility which, in the present approach,
is accomplished by the detailed balance condition. 

Our approach to nonequilibrium thermodynamics assumes that certain
quantities used in equilibrium thermodynamics continues to be well
defined quantities, whereas other quantities,
such as temperature and chemical potential, 
cannot always be assigned to a nonequilibrium system.
One assumes, for instance, that it is possible to
assign an entropy and an energy to the system. According to this assumption, 
to each microscopic chemical state, one associates an energy.
When a reaction occurs, the microscopic chemical state changes causing
an increase or decrease in the energy of the system. This variation
in energy is understood as the energy of activation. In the case of
a system in contact with a heat reservoir, which is the case of the present
approach, the variation in energy is due to the energy exchange with
the reservoir and the rate of reaction is proportional to the
Arrhenius factor. 
The coupling with reservoirs usually is in accordance with the
hypothesis of local equilibrium, that is, the thermodynamic relations
remain valid at a coarse-grained level, as considered, for example, in reference
\cite{qian2016a} for the mesoscopic stochastic formulation of
nonequilibrium thermodynamics.

The present approach assumes that {\it a closed reactive system
will be found in thermodynamic equilibrium when it reaches the steady state},
which amounts to say that the transition rates
associated to the reactions obey the detailed balance condition
in closed system.
It is implied here that the presence of a reaction means that
its reverse is also present.
One way of taking the system out of equilibrium, 
so that the reactions will be unbalanced, 
even in a steady state,
is to place the system in contact with particle reservoirs
in which case the system is open. 
This is what we do in the present approach by representing the contact
with a particle reservoir by a chemical reaction.
Therefore, in addition to the set of ordinary chemical reactions,
another set of chemical reactions
will be considered in order to describe the contact with the particle reservoirs. 

If the open system, in contact with particle reservoirs, reaches equilibrium,
it will be described by the grand canonical Gibbs distribution.
However, the equilibrium will not happen
if the rates of reactions
do not obey detailed balance with respect to the grand canonical distribution.
In this case the system will be in a nonequilibrium situation and
there will be fluxes of particles
between the system and the particle reservoirs, and in general
each reactions will be shifted either to
the products or to the reactants. A flux of entropy from the system to the reservoirs 
will also occur due to the continuous production of entropy. 

We demonstrate that in the nonequilibrium stationary state,
the rate of entropy production is a sum of bilinear terms 
in the affinities and the fluxes of reaction.
In addition, we show that the affinity is written in terms
of the rate constants and the flux of reaction in terms of
the transition rates. 
The derivation of the bilinear form was possible due to
our use of an appropriate form of the transition rates
associated to the chemical reactions and to
the contact of the reaction system with the particle reservoirs.

\section{Microscopic representation}

Our object of study is an open chemical system consisting of molecules 
of several chemical species, or particles of various types,
that react among themselves. The system is in contact with
a heat reservoir and also in contact with several
particle reservoirs, one for each type of particle.  
We assume that the system is described by a continuous
time Markov process defined on a discrete microscopic space of states.
A stochastic trajectory in the microscopic space of states is determined
by the transition rate $W(\eta,\eta')$ from state $\eta'$ to state $\eta$,
a quantity that plays a fundamental role in the present
approach in the sense that a specific system is considered to be
fully characterized when these transition rates are given. 
In other words, all the microscopic processes taking place inside the system,
specifically, the chemical reactions, 
as well as the contact of the system with the reservoirs
are embodied in the transition rates $W(\eta,\eta')$.
Given the transition rates $W(\eta,\eta')$, we set up the master equation
\begin{equation}
\frac{d}{dt}P(\eta) = \sum_{\eta'} \{W(\eta,\eta')P(\eta') - W(\eta',\eta)P(\eta)\},
\label{1}
\end{equation}
which governs the time evolution of the
probability $P(\eta,t)$ of state $\eta$ at time $t$.

For long times the system eventually reaches a stationary state,
meaning that the probability distribution $P(\eta,t)$
approaches a final stationary distribution.
The final stationary state may or may not be an equilibrium state
depending on the transition rates. If the transition rates obey the microscopic
reversibility, that is, if they obey detailed balance with respect to the
final probability distribution, then we say that the system has reached
thermodynamic equilibrium and the equilibrium probability distribution
will be a Gibbs probability distribution.

According to our assumptions, an energy $E(\eta)$ is always associated to the system.
Given the transition rates, this quantity cannot be an arbitrary function but
should be related to the transition rates.
If, for a certain set of values of the parameters defining
the transition rates, these obey detailed balance
with respect to a Gibbs probability distribution, then 
this distribution should involve $E(\eta)$.
That is, in equilibrium, the transition rates
fulfill detailed balance with respect to a Gibbs probability
distribution involving this quantity. 
At this point, however, what we wish to say is that,
from the master equation, it is possible to obtain the
time evolution of the average $U$ of the energy,
\begin{equation}
U = \sum_\eta E(\eta) P(\eta).
\end{equation}
Taking the time derivative of both sides of this equation, and using
the master equation, we immediately find the time evolution of $U$, that is,
\begin{equation}
\frac{dU}{dt} = \sum_{\eta,\eta'}W(\eta,\eta')P(\eta')[E(\eta)-E(\eta')].
\label{5}
\end{equation}
From the master equation we can in fact obtain the time evolution
of any quantity that is an average of a state function. This is the case
of the number of particle of each species.
The average number of particles $N_i(\eta)$ of type $i$ is 
\begin{equation}
N_i = \sum_\eta n_i(\eta) P(\eta),
\end{equation}
where $n_i(\eta)$ stands for the number of particle of state $\eta$.
In an analogous fashion we get from the master equation 
the time evolution of $N_i$, that is, 
\begin{equation}
\frac{dN_i}{dt} = \sum_{\eta,\eta'}W(\eta,\eta')P(\eta')[n_i(\eta)-n_i(\eta')].
\label{6}
\end{equation}

Together with the energy $U$ and the number of particles $N_i$ of each
species, a relevant thermodynamic quantity that characterizes the
system is the entropy. The entropy is not the average of a state function
and is defined by the Gibbs expression
\begin{equation}
S = -k_B \sum_\eta P(\eta)\ln P(\eta),
\label{7}
\end{equation}
assumed to be valid in equilibrium as well as in
nonequilibrium situations, where $k_B$ is the Boltzmann constant.
Its time evolution can be obtained from the master equation and is given
\begin{equation}
\frac{dS}{dt} 
= k_B \sum_{\eta,\eta'}W(\eta,\eta')P(\eta')\ln\frac{P(\eta')}{P(\eta)}.
\label{7a}
\end{equation}
The right-hand side of equation (\ref{7a}) represents the total
variation of entropy, which is understood of consisting of
two parts. One part is the flux of entropy {\it from} the environment,
denoted by $\Phi$, and the other is the rate of production or generation of 
entropy, denoted by $\Pi$. The variation of the entropy of
the system and these two quantities are related by
\cite{prigogine1947,prigogine1950,prigogine1955,groot1962,glansdorff1971}
\begin{equation}
\frac{dS}{dt} = \Pi + \Phi.
\label{9}
\end{equation}
This fundamental relation was advanced by Prigogine
\cite{prigogine1947,prigogine1950,prigogine1955}, 
who wrote it as $dS=dS_i+dS_e$, and
was founded on the ideas of 
De Donder \cite{donder1922,donder1925,donder1927}
and Clausius \cite{clausius1865}
about the ''uncompensated heat''.

To develop a stochastic approach to thermodynamics we need
a microscopic definition of either $\Pi$ or $\Phi$ since the sum
$\Pi+\Phi$ is given by the right-hand side of equation (\ref{7a}).
The definition of the rate of entropy production $\Pi$ should meet 
two conditions: it should be nonnegative and should
vanish in equilibrium, that is, when detailed balance is obeyed. 
This is provided by the Schnakenberg expression \cite{schnakenberg1976}
\begin{equation}
\Pi = k_B \sum_{\eta,\eta'}W(\eta,\eta')P(\eta')
\ln\frac{W(\eta,\eta')P(\eta')}{W(\eta',\eta)P(\eta)},
\label{8}
\end{equation}
which can easily be shown to be semi-positive defined, that is, $\Pi\geq0$. 
The entropy flux $\Phi$ is obtained by replacing
expressions (\ref{8}) and (\ref{7a}) into (\ref{9}). The result is 
\begin{equation}
\Phi = - k_B \sum_{\eta,\eta'}W(\eta,\eta') P(\eta')
\ln\frac{W(\eta,\eta')}{W(\eta',\eta)}.
\label{8a}
\end{equation}

The equations we have introduced in this section define the stochastic
thermodynamics for equilibrium and nonequilibrium systems. However, 
the transition rates were not yet specified.

\section{Transition rates}

Now we wish to set up the transition rates related to the several
chemical reactions occurring inside the system among
$q$ chemical species. The reactions
are described by the chemical equations
\begin{equation}
\sum_{i=1}^q\nu_{ij}^- B_i \rightleftharpoons
\sum_{i=1}^q\nu_{ij}^+ B_i,   \qquad\qquad j=1,2,\ldots,r,
\label{10}
\end{equation}
where $B_i$ denotes the chemical formula of species $i$,
and $\nu_{ij}^- \geq0$ and $\nu_{ij}^+ \geq0$
are the stoichiometric coefficients of the reactants and products, respectively,
and $r$ is the number of reactions.
Equation (\ref{10}) tells us that when the $j$-th reaction occurs from left
to right (forward reaction) then
$\nu_{ij}^-$ molecules of type $i$ disappear and 
$\nu_{ij}^+$ molecules of type $i$ appear so that
the number of molecules of type $i$ varies by 
$\nu_{ij}=\nu_{ij}^+ - \nu_{ij}^-$.
If the reaction occurs from right to left (backward reaction) 
the number of molecules of type $i$ varies by
$-\nu_{ij}=\nu_{ij}^- - \nu_{ij}^+$.
The set of reactions are assumed to be linearly independent, which
means to say that no reaction is a linear combination of the 
others.

The description that we consider here takes into account only the
degrees of freedom related to the variables that specify the chemical
species of each molecule, which we call {\it microscopic chemical state}.
The microscopic chemical state is defined as follows.
In a reaction, we may say that a molecule at position $i$
is transformed into a molecule of distinct type that remains
in the same position $i$. To describe 
this situation we attach a stochastic variable $\eta_i$ at position $i$
that takes values according to the type of molecule present
at position $i$. We adopt the convention that $\eta_i$ takes
the values $1,2,\ldots,q$, according to whether the position $i$
is occupied by a molecule of types $1,2,\ldots,q$, respectively.
If position $i$ is not occupied by any molecule, then $\eta_i$
takes the value zero.
The microscopic chemical state $\eta$ of the whole system is understood as
a vector with components $\eta_i$. 

We assume that the allowed positions, or sites, are finite in number
and form a space structure, that is, a lattice of allowed sites. 
The total number of
sites $N$ of the lattice is proportional to the volume $V$ of the 
recipient and the mean volume $v_c=V/N$ of a cell around a site 
is of the order of the volume of a molecule. In the study of
chemical kinetics it is usual to deal with quantities that are
densities per unit volume. In the present theory, one naturally
deals with densities per site. To get the former density,
it suffices to divide that latter density by $v_c$.

A more complete microscopic description should take into account other degrees
of freedom such as those related to the motion of the molecules
in which case the position $i$ of a molecule should be understood as a dynamical
variable. However, as usually done in the study of chemical kinetics
\cite{decker2016}, we
assume that microscopic chemical degrees of freedom are decoupled
from the mechanical degrees of freedom, or that the coupling
between these two types of degrees of freedom is small.
However, the coupling cannot be entirely avoided because the
energy released or consumed in a chemical reaction is exchanged 
in processes involving the mechanical degrees of freedom such
as the kinetic and potential energies of the molecules.

A chemical reaction described by expression (\ref{10}) can be
understood as the annihilation of a group of particles
and the creation of another group of particles, which
allows to identify the reaction as a transformation
of the state $\eta$ into another state $\eta'$. 
We denote by $R_j^+(\eta',\eta)$ and by $R_j^-(\eta',\eta)$ the
transition rates from $\eta$ to $\eta'$ corresponding to the $j$-th
forward and backward reaction (\ref{10}), respectively. 
To set up these transition rates, we proceed as follows.
We let the system be in contact with a heat reservoir
at a temperature $T$ and assume that the system reaches
the thermodynamic equilibrium. This amounts to say
that detailed balance is fulfilled, that is,
\begin{equation}
R_j^+(\eta,\eta')P^{\,e}(\eta') = R_j^-(\eta',\eta)P^{\,e}(\eta),
\label{14}
\end{equation}
for any pair of states $(\eta,\eta')$ where $P^{\,e}(\eta)$
is the equilibrium Gibbs probability distribution,
\begin{equation}
P^{\,e}(\eta) = \frac1Z e^{-\beta E(\eta)},
\label{15}
\end{equation}
where $E(\eta)$ is the energy of state $\eta$ and $\beta=1/k_BT$.
Therefore, the transition rates of the forward and backward
reactions are connected by the relation
\begin{equation}
\frac{R_j^+(\eta',\eta)}{R_j^-(\eta,\eta')} = e^{-\beta[E(\eta')-E(\eta)]}.
\label{16}
\end{equation}
It should be noted that the right-hand side of this equation
can be regarded as a microscopic Arrhenius factor
\cite{arrhenius1889,moore1965}, the difference
$E(\eta')-E(\eta)$ being the activated energy for the
transition $\eta\to\eta'$. The transition rates we shall consider
are partially defined by this equation, that is, if the forward
transition rate is given then the backward transition rate is
defined by equation (\ref{16}), and vice-versa.

Next we wish to consider the system in
contact with particle reservoirs, one for each type of molecule. 
In this new situation, we assume that the transition rates
$R_j^\sigma(\eta',\eta)$, $\sigma=\pm1$, remains unmodified.
That is, the contact with the particle reservoirs, 
do not modify its form, and equation (\ref{16}) should be
understood as an equation that  defines, or partially defines,
the reaction transition rates. Notice that, equation (\ref{16})
should not be understood as a detailed balance condition because the
equilibrium probability is no longer given by (\ref{15}).

In addition to the transition rates related to the chemical reactions,
we should consider the transitions that describe the contact
with the reservoirs. To find the corresponding transition rates
we consider again the situation in which the
system is found in thermodynamic equilibrium, described by
the following Gibbs probability distribution 
\begin{equation}
P^{\,e}(\eta) = \frac1\Xi e^{-\beta E(\eta)+\beta\sum_i \mu_i n_i(\eta)},
\label{20}
\end{equation}
where $n_i(\eta)$ is the number of molecules of type $i$ in state $\eta$
and $\mu_i$ is the chemical potential associated to reservoir $i$.
Denoting by $C_i^+(\eta',\eta)$ and $C_i^-(\eta',\eta)$ the
transition rates corresponding to the addition and removal of one
particle of type $i$, respectively, then these rates obey the relation
\begin{equation}
\frac{C_i^+(\eta',\eta)}{C_i^-(\eta,\eta')}
= e^{-\beta[E(\eta')-E(\eta)] + \beta\mu_i [n_i(\eta')-n_i(\eta)]}.
\label{19}
\end{equation}
We are considering that just one molecule is added to or removed from the
system so that in this equation, $n_i(\eta')-n_i(\eta)=+1$. 
Equation (\ref{19}) is assumed to be an equation that defines,
or partially defines, the transition rates associated to the
contact with the reservoirs. The motivation for this definition
is the following. If the system has no reaction, that is,
if the only transition rates are $C_i^\sigma(\eta,\eta')$, $\sigma=\pm1$,
then in the steady state the system
will be found in equilibrium with the distribution (\ref{20})
because (\ref{19}) is identified, in this case, with detailed balance
with respect to (\ref{20}).

The stochastic approach to equilibrium and nonequilibrium
thermodynamics of chemical reactions
that we are considering here is founded on the master equation (\ref{1})
with transition rates $W(\eta,\eta')$ given by
\begin{equation}
W(\eta,\eta') = \sum_{j=1}^r \sum_{\sigma=\pm1} R_j^{\,\sigma}(\eta,\eta')
+ \sum_{i=1}^q \sum_{\sigma=\pm1} C_i^{\,\sigma}(\eta,\eta').
\label{41}
\end{equation}
We remark that the matrices $R_j^{\,\sigma}$ and $C_i^{\,\sigma}$
are disjoint, that is,
if the entry $(\eta,\eta')$ of one of them is nonzero, then
the same entry of any other vanishes. 
In other words, depending on the states $\eta$ and $\eta'$,
the transition rate $W(\eta,\eta')$  
is either one of the reaction transition rates $R_j^{\,\sigma}(\eta,\eta')$
or one of the contact transition rates $C_i^{\,\sigma}(\eta,\eta')$,
which obey equations (\ref{16}) and (\ref{19}), respectively. 

Given these rates
we may ask whether they obey detailed balance with respect
to the equilibrium probability distribution (\ref{20}).
By construction, the rates $C_i^{\,\sigma}$ indeed obey it. 
But in general the rates $R_j^{\,\sigma}$ do not. The
detailed balance condition for $R_j^{\,\sigma}$, with respect
to the equilibrium distribution (\ref{20}), is
\begin{equation}
\frac{R_j^+(\eta',\eta)}{R_j^-(\eta,\eta')}
= e^{-\beta [E(\eta')-E(\eta)]+\beta\sum_i \mu_i [n_i(\eta')-n_i(\eta)]}.
\label{17}
\end{equation}
But the left hand side should be given by equation (\ref{16}).
A comparison between equations (\ref{17}) and (\ref{16}) leads us 
to conclude that $R_j^{\,\sigma}$ does not obey detailed balance
unless the summation on the exponent on the right-hand side of
equation (\ref{17}) vanish. Taking into account that
$n_i(\eta')-n_i(\eta) = \nu_{ij}$,
the summation on the exponent vanishes if
\begin{equation}
\sum_i \mu_i \nu_{ij} = 0,
\label{22}
\end{equation}
which is the well known equilibrium condition for a system consisting
of chemical reactions \cite{callen1960,reichl1980,kondepudi1998,oliveira2013}.
If the chemical potentials $\mu_i$
fulfill equation (\ref{22}) for each reaction $j=1,\ldots,r$, then, when
the system reaches the stationary state, it will be found in equilibrium
and described by the Gibbs probability distribution (\ref{20}).
Otherwise, the system will not reach equilibrium and will be found
in a nonequilibrium stationary state.

\section{Nonequilibrium regime}

When the chemical potentials do not obey condition (\ref{22}),
the system will reach a nonequilibrium stationary state
because detailed balance condition is not fulfilled,
and the system cannot be in equilibrium. At least one reaction
is shifted either to the right or to the left, that is,
either the products are  
being created and the reactants being annihilated (forward reaction)
or the reactants are being created and the products being annihilated 
(backward reaction). In the stationary nonequilibrium state,
entropy is continuously being produced and the rate of entropy
production equals the flux of entropy.  Some type of particles
are being created and others annihilated, given rise to fluxes 
of particles either to the system or from the system. The 
set of reactions may be exothermic, in which case the chemical
work is transformed into heat that leaves the system, or 
endothermic, in which case the heat from the outside
is transformed into chemical work.

A nonequilibrium situation is characterized by the existence of
fluxes of distinct types such as the energy flux, the particle flux
and the entropy flux. If a certain quantity is a conserved quantity
then its time variation should be equal to the input flux.
This is the case of energy. If we denote
by $\Phi_u$ the flux of energy, that is, the energy per unit time,
{\it received} by the system from the reservoir then
\begin{equation}
\frac{dU}{dt} = \Phi_u.
\label{76}
\end{equation}
Comparing this equation with (\ref{5}), we find the following expression for
the energy flux
\begin{equation}
\Phi_u = \sum_{\eta,\eta'} W(\eta,\eta') P(\eta')[E(\eta)-E(\eta')].
\label{28}
\end{equation}

In addition to the flux of heat, the system may also be subject to the
flux of particles. Taking into account that the flux of particles is a 
consequence of the contact with the particle reservoirs, which are
described by the transition rates $C_i^{\,\sigma}(\eta,\eta')$, it
follows that the flux of particles $\Phi_i$ of type $i$ is expressed by
\begin{equation}
\Phi_i = \sum_{\eta,\eta'} \sum_{\sigma=\pm1} C_i^{\,\sigma}(\eta,\eta')
P(\eta')[n_i(\eta)-n_i(\eta')],
\label{27}
\end{equation}
which can be written as
\begin{equation}
\Phi_i = \sum_{\eta,\eta'}
[C_i^+(\eta,\eta') - C_i^-(\eta,\eta')] P(\eta').
\label{27a}
\end{equation}
In chemical reactions, particles can be created or annihilated.
In this sense the number of particles of a certain species may not be a
conserved quantity and as a consequence its time variation may not be equal to flux
of particle of this species. Accordingly, we write
\cite{prigogine1947,prigogine1950,prigogine1955,groot1962}
\begin{equation}
\frac{dN_i}{dt} = \Gamma_i + \Phi_i,
\label{31}
\end{equation}
where $\Gamma_i$ is interpreted as the rate in which particles of type $i$ are being
created ($\Gamma_i>0$) or annihilated ($\Gamma_i<0$) inside the system.
Comparing this equation with (\ref{6}) and taking into account equation
(\ref{27}), we see that
\begin{equation}
\Gamma_i = \sum_{\eta,\eta'} \sum_{j=1}^r \sum_{\sigma=\pm1}
R_j^{\,\sigma}(\eta,\eta') P(\eta')[n_i(\eta)-n_i(\eta')].
\label{37}
\end{equation}
Bearing in mind that in the $j$-th forward reaction,
the number of particles of type $i$ varies by $\nu_{ij}$
and in the backward by $-\nu_{ij}$, this equation can be written as
\begin{equation}
\Gamma_i = \sum_{j=1}^r \nu_{ij} X_j,
\label{37a}
\end{equation}
where
\begin{equation}
X_j = \sum_{\eta,\eta'} [R_j^+(\eta,\eta') - R_j^-(\eta,\eta')] P(\eta')
\label{37b}
\end{equation}
is the flux of $j$-th reaction.
If $X_j>0$, the $j$-th reaction is shifted to the right, toward the
products. If $X_j<0$, it is shifted to the left, toward the reactants.

As we have seen, the time variation of the entropy of the system is
\begin{equation}
\frac{dS}{dt} = \Pi + \Phi,
\end{equation}
which means to say that entropy is also a nonconserved quantity,
but differently from the number of particles, it cannot decrease
because $\Pi\geq0$, which is the expression of the second law
of thermodynamics. The replacement of (\ref{41}) into equation
(\ref{8a}), furnishes an expression for the 
entropy flux $\Phi$ in terms of the reaction transition rates
$R_j^{\,\sigma}$ and contact transition rates $C_i^{\,\sigma}$.
When the resulting expression is compared with the right-hand sides
of equations (\ref{28}) and (\ref{27}), we see that
the entropy flux $\Phi$ is related to
the energy flux $\Phi_u$ and particle fluxes $\Phi_i$ by 
\begin{equation}
\Phi = \frac1T (\Phi_u - \sum_{i=1}^q \mu_i \Phi_i).
\label{34}
\end{equation}

The flux of heat from the thermal reservoir is defined as the flux of energy
plus the rate of chemical work performed by the system, that is,
\begin{equation}
\Phi_q = \Phi_u - \sum_{i=1}^q \mu_i\Phi_i.
\label{29}
\end{equation}
From this expression we may conclude
that
\begin{equation}
\Phi = \frac1T \Phi_q,
\label{33}
\end{equation}
that is, the flux of entropy equals the heat flux divided by
the temperature, in accordance with Clausius \cite{clausius1865}.

Let us consider the nonequilibrium stationary state.
In this regime $dU/dt=0$, implying $\Phi_u=0$ so that
equation (\ref{34}) reduces to 
\begin{equation}
\Phi = - \frac1T \sum_{i=1}^q \mu_i\Phi_i  = \frac1T \sum_{i=1}^q \mu_i\Gamma_i,
\end{equation}
where we have used the result $\Phi_i=-\Gamma_i$ because $dN_i/dt=0$. 
But in the stationary state $dS/dt=0$, implying $\Pi=-\Phi$
and as a consequence  
\begin{equation}
\Pi = - \frac1T \sum_{i=1}^q \mu_i\Gamma_i.
\end{equation}
Taking into account the result (\ref{37a}) and defining the 
affinity ${\cal A}_j^D$ by 
\cite{donder1927,prigogine1947,prigogine1950,prigogine1955,groot1962,glansdorff1971}
\begin{equation}
{\cal A}_j^{D} = - \sum_{i=1}^q \mu_i \nu_{ij},
\label{45a}
\end{equation}
then the rate of entropy production can be written in the bilinear form
\cite{prigogine1947,prigogine1950,prigogine1955,groot1962,glansdorff1971}
\begin{equation}
\Pi = \frac1T \sum_{j=1}^r {\cal A}_j^D X_j.
\end{equation}
The concept of affinity was introduced by De Donder 
\cite{donder1922,donder1925} whereas the 
the bilinear form for the production of entropy was
advanced by Prigogine \cite{prigogine1947,prigogine1955}.
Here we find it more convenient to defined the affinity as
the expression (\ref{45a}) divided by the temperature,
\begin{equation}
{\cal A}_j = - \frac1T \sum_{i=1}^q \mu_i \nu_{ij},
\label{45}
\end{equation}
so that
\begin{equation}
\Pi = \sum_{j=1}^r {\cal A}_j X_j.
\end{equation}
Notice that, in equilibrium, not only ${\cal A}_j=0$ but also $X_j=0$.

\section{Number of particles representation}

Let us apply the present approach to the case in which
the energy $E(\eta)$ depends on $\eta$ only through the
number of particle $n_i(\eta)$ of each species. 
We use the notation $E(n)$ where $n$ is a vector with components
$n_i$, $i=1,2,\ldots,q$.
It is then possible to assume that the reaction transition rates
$R_j^{\,\sigma}(\eta,\eta')$ and the contact transition rates
$C_i(\eta,\eta')$ depend on $\eta$ and $\eta'$ only through
the numbers of particle of each species. The description of the
system can thus be made in terms of the stochastic variables
$n_1,n_2,\ldots,n_q$.
Assuming that $P(\eta)$ depends on $\eta$ only trough $n_i(\eta)$
then the probability $\bar{P}(n)$ of $n$ must be related to $P(\eta)$ by
$\bar{P}(n)=A(n)P(\eta)$ where 
\begin{equation}
A(n) = \frac{N!}{n_0!n_1!\ldots n_q!},
\label{39}
\end{equation}
and $n_0=N-(n_1+\ldots+n_q)$ is the number of empty sites. 
The equilibrium probability distribution in the new representation is thus
\begin{equation}
\bar{P}^{\,e}(n) = \frac{A(n)}{\Xi} e^{-\beta E(n) + \beta\sum_i\mu_i n_i}.
\end{equation}
Analogously, the transition rate $\bar{W}(n,n')$ from $n'$ to $n$
is related to the transition rate $W(\eta,\eta')$ of the original
representation by $\bar{W}(n,n')=A(n)W(\eta,\eta')$.
In the new representation, the master equation (\ref{1}) becomes
\begin{equation}
\frac{d}{dt}\bar{P}(n) = \sum_{n'} \{\bar{W}(n,n')\bar{P}(n')-\bar{W}(n',n)\bar{P}(n)\}.
\label{55}
\end{equation}

Let us write the entropy, given by (\ref{7}), in the new representation,
\begin{equation}
S = -k_B \sum_n \bar{P}(n)\ln \frac{\bar{P}(n)}{A(n)}.
\end{equation}
The expression for the production of entropy (\ref{8}) and entropy flux
(\ref{8a}) in the new representation are
\begin{equation}
\Pi = k_B \sum_{n,n'}\bar{W}(n,n')\bar{P}(n')
\ln\frac{\bar{W}(n,n')\bar{P}(n')}{\bar{W}(n',n)\bar{P}(n)},
\end{equation} 
\begin{equation}
\Phi = - k_B \sum_{n,n'}\bar{W}(n,n') \bar{P}(n')
\ln\frac{\bar{W}(n,n')A(n')}{\bar{W}(n',n)A(n)}.
\label{61}
\end{equation}
It should be noted that the production of entropy in the new
representation has the same form of the original representation $\eta$,
although that is not true for the entropy and flux of entropy. 

The transition rate $\bar{W}(n',n)$ is either the transition
rate related to one of the reactions (\ref{10}) or the transition rate related
to contact with a particle reservoir. 
We denote the former by $R_j^{\,\sigma}(n)$ and the latter by
$C_i^{\,\sigma}(n)$. More precisely, $R_j^+(n)$ is the
transition rate from $n$ to $n^j$, where $n^j$ the state obtained from $n$ by
the action of the forward reaction $j$, which amounts to say that
\begin{equation}
n_i^j-n_i=\nu_{ij}^+ - \nu_{ij}^- =\nu_{ij},
\end{equation}
whereas
$C_j^+(n)$ is the transition rate from $n$ to $n^i$, where
$n^i$ stands for the state $n$ with one more particle of type $i$
so that $n_i^i-n_i=1$. The transition rates $R_j^-(n)$
and $C_j^-(n)$ are defined similarly.
These transition rates obey the equations
\begin{equation}
\frac{R_j^+(n)}{R_j^-(n^j)} = \frac{A(n^j)}{A(n)} e^{-\beta[E(n^j)-E(n)]},
\label{16a}
\end{equation}
and
\begin{equation}
\frac{C_i^+(n)}{C_i^-(n^i)} = \frac{A(n^i)}{A(n)}
e^{-\beta[E(n^i)-E(n)] + \beta\mu_i},
\label{19a}
\end{equation}
which come from equations (\ref{16}) and (\ref{19}), respectively.

The expression for the flux of particle becomes
\begin{equation}
\Phi_i = \sum_{n} [C_i^+(n) - C_i^-(n)]\bar{P}(n)
= \langle C_i^+\rangle - \langle C_i^-\rangle,
\label{43}
\end{equation}
whereas the expression for the flux of the reaction $X_j$ is
\begin{equation}
X_j = \sum_{n} [R_j^+(n) - R_j^-(n)]\bar{P}(n)
= \langle R_j^+\rangle - \langle R_j^-\rangle.
\label{44}
\end{equation}

\section{Rates of the chemical kinetics}

To proceed further on we need to know how $E(n)$ depends on $n$.
Here we consider the simplest case in which the energy
depends linearly on the number of particles, that is,
\begin{equation}
E(n) = \sum_{i=1}^q \varepsilon_i n_i,
\label{47}
\end{equation}
where $\varepsilon_i$ is the energy associated to a particle of type $i$.
The variation in energy associated to the $j$ forward reaction is thus
\begin{equation}
E(n^j) - E(n) = \sum_{i=1}^q \varepsilon_i \nu_{ij}.
\label{48}
\end{equation}

The knowledge of the dependence of $E(n)$ on $n$ is not sufficient to determine
the transition rates $R_j^{\,\sigma}$ and $C_i^{\,\sigma}$
since only their ratios are known in terms of $E(n)$,
as follows from equations (\ref{16a}) and (\ref{19a}). 
There is thus a great deal of freedom in the establishment of
the transition rates. As we will see below, we will set up transitions
rates that are in accordance with those used in the area of chemical kinetics,
also known as transitions rates coming from the laws of mass action.

Instead of the expression (\ref{16a}) for $A(n)$,
we use the following expression
\begin{equation}
A(n) = \frac{N^{n_1+n_2+\ldots+n_q}}{n_1!n_2!\ldots n_q!},
\label{46}
\end{equation}
which is obtained from (\ref{16a}) by assuming that the number
of empty sites $n_0$ is great enough.
Using expression (\ref{46}) for $A(n)$ and (\ref{47}) for $E(n)$,
then equation (\ref{16a}) is written as
\begin{equation}
\frac{R_j^+(n)}{R_j^-(n^j)}
= \frac{A(n^j)}{A(n)} \prod_{i=1}^q e^{-\beta\varepsilon_i\nu_{ij}}
= \prod_{i=1}^q \frac{n_i!}{n_i^j!} \left(Ne^{-\beta\varepsilon_i}\right)^{\nu_{ij}}. 
\label{63}
\end{equation}
A solution, which is in agreement with the laws of mass action, is
\begin{equation}
R_j^+(n) = k_j^+ N \prod_{i=1}^q \frac{n_i!}{(n_i-\nu_{ij}^-)! N^{\nu_{ij}^-}},
\label{60a}
\end{equation}
\begin{equation}
R_j^-(n) = k_j^+ N \prod_{i=1}^q \frac{n_i!}{(n_i-\nu_{ij}^+)! N^{\nu_{ij}^+}},
\label{60b}
\end{equation}
where the constants of reaction $k_j^+$ and $k_j^-$ must obey relation
\begin{equation}
\frac{k_j^+}{k_j^-}=\prod_{i=1}^q e^{-\beta\varepsilon_i \nu_{ij}}.
\label{35a}
\end{equation}
Since $n_i$ is much larger than the stoichiometric coefficients,
we may write
\begin{equation}
R_j^+(n) = k_j^+ N \prod_{i=1}^q \left( \frac{n_i}{N} \right)^{\nu_{ij}^-},
\label{60c}
\end{equation}
\begin{equation}
R_j^-(n) = k_j^- N \prod_{i=1}^q \left( \frac{n_i}{N} \right)^{\nu_{ij}^+},
\label{60d}
\end{equation}
which are in accordance with the law of mass action \cite{yeremin1979}.

Similarly, using expression (\ref{46}) for $A(n)$ and the result (\ref{47})
for $E(n)$, then equation (\ref{19a}) is written as
\begin{equation}
\frac{C_i^+(n)}{C_i^-(n^i)} = \frac{N}{n_i+1}
e^{-\beta\varepsilon_i + \beta\mu_i}.
\end{equation}
A solution is 
\begin{equation}
C_i^+(n) = c_i^+ N,
\label{60e}
\end{equation}
\begin{equation}
C_i^-(n) =  c_i^- n_i,
\label{60f}
\end{equation}
where $c_i^+$ and $c_i^-$ should obey relation
\begin{equation}
\frac{c_i^+}{c_i^-} = e^{-\beta\varepsilon_i + \beta\mu_i}.
\label{36a}
\end{equation}

For convenience we define $x_i=N_i/N$ and the functions
\begin{equation} 
w_j^+(x) = k_j^+ \prod_{i=1}^q x_i^{\,\nu_{ij}^-},
\qquad\qquad
w_j^-(x) = k_j^- \prod_{i=1}^q x_i^{\,\nu_{ij}^+},
\label{51}
\end{equation}
and
\begin{equation}
v_i^+(x) = c_i^+,
\qquad\qquad
v_i^-(x) =  c_i^- x_i.
\end{equation}
In terms of these functions, the transition rates are
$R_j^{\,\sigma} = N w_j^{\,\sigma}$ and $C_i^{\,\sigma} = N v_i^{\,\sigma}$.

\section{Steady state and entropy production}

A solution of the master equation (\ref{55}) with the transition
rates (\ref{60c}), (\ref{60d}), (\ref{60e}), and (\ref{60f})
can easily be obtained in the regime of large $N$.
In this regime, the distribution $\rho(x)=NP(n)$
of the variables $x_i=n_i/N$ will be peaked around the averages
$\bar{x}_i=\langle x_i\rangle$. In fact the distribution will be
a multivariate Gaussian distribution with variances proportional to $N$.
Therefore, in the limit $N\to\infty$, the average $\langle f(x)\rangle$ of a
function of $x$ may be replaced by $f(\bar{x})$.
Using this result in equation (\ref{43}), we see that the
flux of particles $\phi_i=\Phi_i/N$ per site is
\begin{equation}
\phi_i = v_i^+(\bar{x}) - v_i^-(\bar{x}).
\label{53}
\end{equation}
Using the same result in equation (\ref{44}), the flux of reaction per site
$\chi_j=X_j/N$ is written as
\begin{equation}
\chi_j = w_j^+(\bar{x}) - w_j^-(\bar{x}).
\label{53a}
\end{equation}
Equation (\ref{31}), that gives the time evolution of $\langle n_i\rangle$,
is thus written as
\begin{equation}
\frac{d\bar{x}_i}{dt} = \gamma_i + \phi_i,
\label{40}
\end{equation}
where
\begin{equation}
\gamma_i = \sum_{j=1}^r \nu_{ij}\chi_j.
\label{40z}
\end{equation}
The equation (\ref{40}) constitute a set of closed equations for $\bar{x}$.

In the stationary state we may solve for $\bar{x}$, and obtain
the rate of entropy production per site ${\sf P}=\Pi/N$, given by
\begin{equation}
{\sf P} = \sum_{j=1}^r {\cal A}_j \chi_j.
\label{89}
\end{equation}

It is worth writing the affinities in terms of the constants of reaction,
\begin{equation}
{\cal A}_j = k_B 
\left(\ln\frac{k_j^+}{k_j^-} - \sum_{i=1}^q \nu_{ij}\ln\frac{c_i^+}{c_i^-}\right),
\end{equation}
obtained from its definition (\ref{45}) and from the relations
(\ref{35a}) and (\ref{36a}).
Therefore, we may write the production of entropy as 
\begin{equation}
{\sf P} = k_B  \sum_{j=1}^r 
\left(\ln\frac{k_j^+}{k_j^-} - \sum_{i=1}^q \nu_{ij}\ln\frac{c_i^+}{c_i^-}\right)
(w_j^+ - w_j^-).
\end{equation}


A simplification on the approach just presented can be obtained by
considering that the rate constants $c_i^+$ and $c_i^-$
related to the contact with the reservoirs are large enough.
Strictly speaking we will take the limit
$c_i^+\to\infty$ and $c_i^-\to\infty$ with the ratio
\begin{equation}
\frac{c_i^+}{c_i^-}=\zeta_i
\end{equation}
finite. According to equation (\ref{36a})
this ratio is $\zeta_i=e^{-\beta\varepsilon_i + \beta\mu_i}$, which
we call the activity related to particle of type $i$, a concept
introduced by Lewis \cite{lewis1923}.

In the present approach it is not necessary that the system 
exchanges particles of all types. We thus suppose that the
system is closed to particles of type $i=1,\ldots,q'$ and
that it is in contact with reservoirs corresponding to
particles of type $k=q'+1,\ldots,q$, so that $q-q'$ is the
number of particle reservoirs. Thus for particles
of type $i=1,\ldots,q'$, the flux $\phi_i$ vanishes identically.
Thus equation (\ref{40}) is split into two types of equations
\begin{equation}
\frac{d\bar{x}_i}{dt} = \sum_{j=1}^r \nu_{ij}\chi_j,
\qquad\qquad i=1,2,\ldots,q'
\label{40a}
\end{equation}
\begin{equation}
\frac{d\bar{x}_k}{dt} = \sum_{j=1}^r \nu_{kj}\chi_j + \phi_k,
\qquad\qquad k=q'+1,\ldots,q.
\label{40b}
\end{equation}

Now, for the second set of species the flux of particles is
\begin{equation}
\phi_k = c_k^+ - c_k^-\bar{x}_k = c_k^-(\zeta_k - \bar{x}_k).
\end{equation}
If $c_k^-$ is large enough, equation (\ref{40}) will be dominated
by this term so that $\bar{x}_k$ reaches very rapidly the value $\zeta_k$.
Therefore, for the second set of species we may set 
\begin{equation}
\bar{x}_k = \zeta_k
\qquad\qquad k=q'+1,\ldots,q,
\end{equation}
and plug it in the right-hand side of equation (\ref{40a}).
This equation is then solved for $\bar{x}_i$, $i=1,\ldots,q'$.

In the stationary state, the entropy production will be
given by 
\begin{equation}
{\sf P} = \sum_{j=1}^r {\cal A}_j (w_j^+ - w_j^-) = \sum_{j=1}^r {\cal A}_j \chi_j, 
\label{69a}
\end{equation}
with the affinity given by
\begin{equation}
{\cal A}_j = k_B \left(\ln\frac{k_j^+}{k_j^-}
- \sum_{k=q'+1}^q \nu_{kj}\ln\zeta_k \right).
\label{69}
\end{equation}

An alternative form to calculate the rate of entropy production is
\begin{equation}
{\sf P} = \sum_{j=1}^r {\cal A}_j \chi_j
- k_B\sum_{i=1}^{q'} \gamma_i \ln \bar{x}_i,
\label{89a}
\end{equation}
which follows from the result that, in this equation,
$\gamma_i=0$ in the stationary state, so that equation (\ref{89a})
becomes identical to equation (\ref{69a}). Using the definition (\ref{40z})
for $\gamma_i$, we get 
\begin{equation}
{\sf P} = \sum_{j=1}^r \chi_j \left( {\cal A}_j  - k_B\sum_{i=1}^{q'} 
 \nu_{ij}\ln \bar{x}_i \right).
\label{89b}
\end{equation}
Now, from (\ref{51}) and (\ref{69}), we find
\begin{equation} 
{\cal A}_j  = k_B \left(
\ln \frac{w_j^+}{w_j^-} + \sum_{i=1}^{q'}  \nu_{ij} \ln \bar{x}_i \right).
\end{equation}
Replacing this result into (\ref{89b}), 
we reach the alternative but equivalent form for the rate
of entropy production \cite{kondepudi1998}
\begin{equation}
{\sf P} = k_B\sum_{j=1}^r (w_j^+ - w_j^-) \ln \frac{w_j^+}{w_j^-}.
\label{69b}
\end{equation}

Although both equations (\ref{69a}) and (\ref{69b}) give the
entropy at the stationary state, they are conceptually distinct.
Expression (\ref{69a}) is a sum of terms, each one being 
a product of a {\it flux} and a {\it thermodynamic force},
or in the present case, a flux of reaction, $\chi_j$,
and an affinity, ${\cal A}_j$.
It should be remarked that the first is a thermodynamic density
and the second, a thermodynamic field, usually called intensive 
variable. In addition, expression (\ref{69a}) is suited 
for an Onsager coefficients \cite{onsager1931},
which is obtained by expanding
$\chi_j$ in terms of ${\cal A}_j$.

If necessary, the fluxes of particles $\phi_k$ can be computed
from the fluxes of reactions $\chi_j$, in the stationary state, by
\begin{equation}
\phi_k = - \sum_{j=1}^r \nu_{kj}\chi_j,
\label{70}
\qquad\qquad k=q'+1,\ldots, q.
\end{equation}

\section{Applications}

In the following we apply the present approach 
to known models of reactive systems.
The models are defined by $r$ reactions of type (\ref{10})
involving $q$ types of particles. The system is close to
particles of type $i=1,\ldots,q'$, and open to particles
of type $k=q'+1,\ldots,q$, and the system is contact
only with reservoirs $q-q'$ reservoirs, only.
According to the formalism that we have just developed
in the second part of the previous section, we may set
\begin{equation}
x_k = \zeta_k,
\qquad\qquad k=q'+1,\ldots,q.
\end{equation}
The evolution equation for the $q'$ densities that may vary is given by
(\ref{40a}), that is,
\begin{equation}
\frac{dx_i}{dt} = \sum_{j=1}^r \nu_{ij}\chi_j,
\qquad\qquad i=1,2,\ldots,q',
\label{50}
\end{equation}
where $\chi_j=w_j^+ - w_j^-$, and $w_j^{\,\sigma}$ 
are the transition rates per site, given by (\ref{51}),
with $\zeta_k$ replacing $x_k$.
Here we are dropping the bar over $x$.

In each case we consider as parameters of the model
the rate constants $k_j^+$ and $k_j^-$ and the activities 
$\zeta_k$. From these quantities we determine the affinities
${\cal A}_j$ by the use of equation (\ref{69}), that is,
\begin{equation}
{\cal A}_j = \ln\frac{k_j^+}{k_j^-}
- \sum_{k=q'+1}^q \nu_{kj}\ln\zeta_k,
\label{65}
\end{equation}
where we have set $k_B=1$.
Equation (\ref{50}) is solved and the densities $x_i$
are determined at the stationary state, which amounts to solve
the equation
\begin{equation}
\sum_{j=1}^r \nu_{ij}\chi_j = 0,
\qquad\qquad i=1,2,\ldots,q'.
\label{72}
\end{equation}
From $x_i$ we may determine the fluxes of reactions $\chi_j$, the fluxes
of particles $\phi_i$, and the rate of entropy production ${\sf P}$
by equation (\ref{69a}), that is,
\begin{equation}
{\sf P} = \sum_{j=1}^r {\cal A}_j \chi_j.
\end{equation}

In the following we apply the approach we have developed to
the case of the first and second Schl\"ogl models 
\cite{schlogl1972}. The production of entropy of the second
model has been determined by several authors
\cite{gaspar2004,vellela2009,endres2015,endres2017}
by means of formula (\ref{69b}). 

\subsection{First Schl\"ogl model}

We start with the case of a chemical
system with two reactions and three types of particles, known as
the first Schl\"ogl model. The reactions are
\begin{equation}
X + A \rightleftharpoons 2X, \qquad\qquad X \rightleftharpoons B,
\end{equation}
and the system is in contact with reservoirs of particle of type A and B, only.
We denote by $x$, $y$, and $z$ the densities of $X$, $A$, and $B$, respectively,
and by $a$ and $b$ the activities of $A$ and $B$, respectively. Then
\begin{equation}
y = a, \qquad\qquad z=b ,
\end{equation}
and 
\begin{equation}
\chi_1 = k_1^+ a x  - k_1^- x^2, \qquad\qquad \chi_2 = k_2^+ x - k_2^- b.
\end{equation}

Equation (\ref{50}), which gives the time evolution of $x$, becomes
\begin{equation}
\frac{dx}{dt} = \chi_1 - \chi_2.
\end{equation}
In the stationary state
\begin{equation}
\chi_1 - \chi_2 = 0.
\end{equation}
Solving this equation for $x$ we find
\begin{equation}
x = \frac{1}{2k_1^-}\left\{k_1^+ a -k_2^+ + 
[(k_1^+ a -k_2^+ )^2 +4 k_1^-k_2^- b]^{1/2} \right\}.
\end{equation}

The affinities are
\begin{equation}
{\cal A}_1 = \ln\frac{k_1^+}{k_1^-} + \ln a,
\label{72a}
\end{equation}
\begin{equation}
{\cal A}_2 = \ln\frac{k_2^+}{k_2^-} - \ln b,
\label{72b}
\end{equation}
and in the stationary state the rate of entropy production is
\begin{equation}
{\sf P} = {\cal A}_1 \chi_1 + {\cal A}_2 \chi_2.
\label{72c}
\end{equation}
From the solution for $x$ we obtain $\chi_1$ and $\chi_2$ and $\Pi$.
The fluxes of particle B and C will be $\phi_2=\chi_1$ and $\phi_3=-\chi_1$.

\begin{figure}
\epsfig{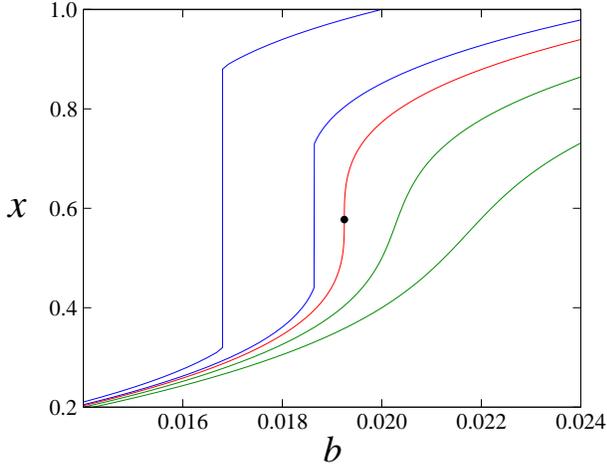}
\caption{Density $x$ as a function of $b$ for $a=0.18$,
0.175, 0.1732, 0.17, and 0.165, from left to right;
and the following values of the rate constants: 
$k_1^+=1$, $k_1^-=0.1$, $k_2^+=0.1$, and $k_2^-=1$.
The full circle represents the critical point and the
vertical straight lines represent discontinuous phase
transitions.}
\label{apsv}

\end{figure}
\begin{figure}
\epsfig{file=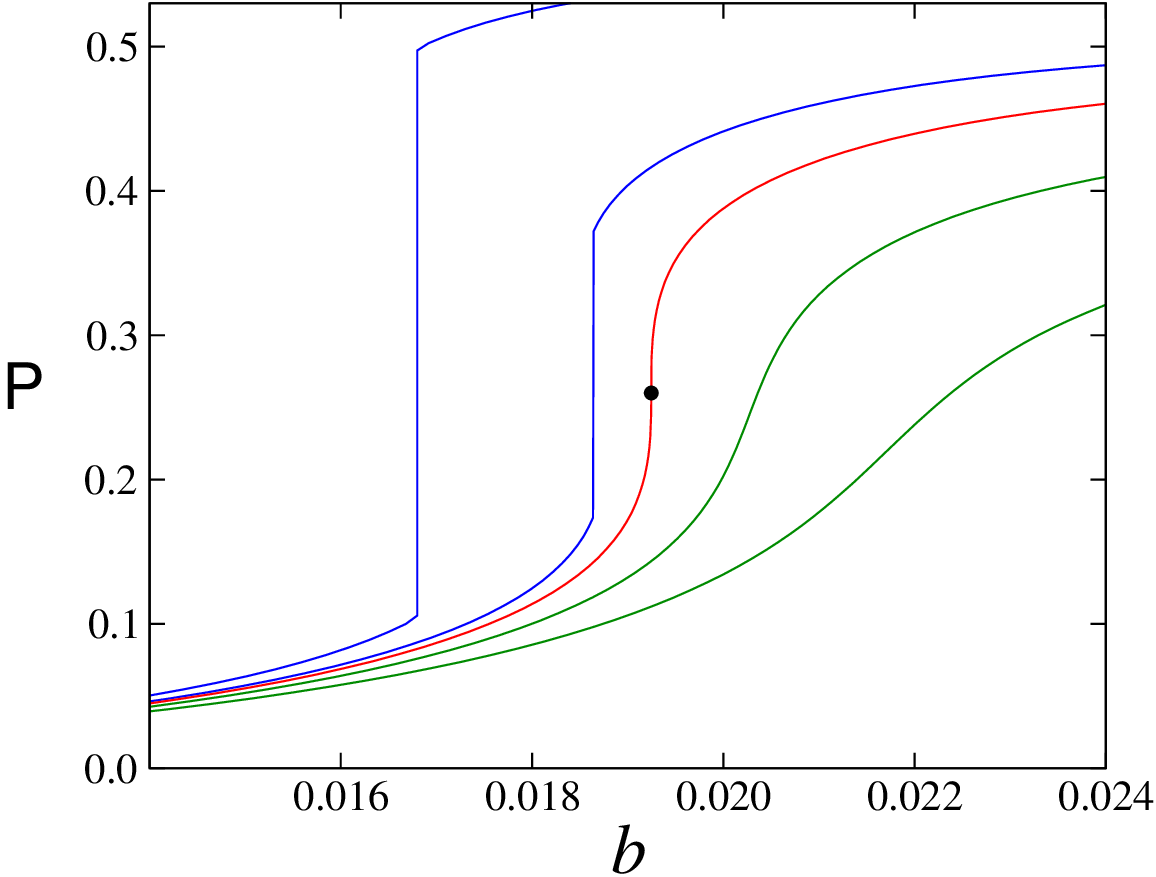,width=8cm}
\caption{Rate of entropy production ${\sf P}$
as a function of $b$. 
The parameters are the same as those of figure \ref{apsv}.}
\label{pe}
\end{figure}

\subsection{Second Schl\"ogl model}

The reactions of the second Schl\"ogl model are
\begin{equation}
2X + A \rightleftharpoons 3X, \qquad\qquad X \rightleftharpoons B,
\end{equation}
and again the system is in contact with reservoirs of particle of type A and B, only.
Again, we denote by $x$, $y$, and $z$ the densities of $X$, $A$, and $B$,
respectively, and by $a$ and $b$ the activities of $A$ and $B$, respectively. Then
\begin{equation}
y = a, \qquad\qquad z=b,
\end{equation}
and 
\begin{equation}
\chi_1 = k_1^+ a x^2  - k_1^- x^3, \qquad\qquad \chi_2 = k_2^+ x - k_2^- b.
\end{equation}

Equation (\ref{50}), which gives the time evolution of $x$, becomes
\begin{equation}
\frac{dx}{dt} = \chi_1 - \chi_2.
\label{50x}
\end{equation}
In the stationary state
\begin{equation}
\chi_1 - \chi_2 = 0,
\end{equation}
which is equivalent to
\begin{equation}
k_1^- x^3 - k_1^+ a x^2   + k_2^+ x - k_2^- b = 0,
\label{80}
\end{equation}
and the density $x$ of particles of type $X$ is the root of this equation.

The affinities ${\cal A}_1$ and ${\cal A}_2$ have the same form
as those of the first model, given by equations (\ref{72a}) and (\ref{72b}).
and the rate of entropy production is
\begin{equation}
{\sf P} = {\cal A}_1 \chi_1 + {\cal A}_2 \chi_2.
\end{equation}
Taking into account that $\chi_1=\chi_2$ we may write
\begin{equation}
{\sf P} = {\cal A} \,\chi,
\end{equation}
where ${\cal A}={\cal A}_1+{\cal A}_2$ and $\chi=\chi_1=\chi_2$.

Solving equation (\ref{50x}), for a given initial condition,
and taking the limit $t\to\infty$, the final
value of $x(t)$ will be a solution of (\ref{80}).
For a given set of the parameters, equation (\ref{80}) may present
a single solution. In this case,
the final value of $x(t)$ will be this single solution no matter what
the initial condition is. For another set of the parameters,
equation (\ref{80}) may have three solutions
and the final value of $x(t)$ will depend on the initial condition.
In this case, 
we have arbitrarily chosen as the initial condition the value of $x$
at the inflexion point when 
the solutions of (\ref{80}) is plotted as a function of $b$,
Under this condition, final value of $x(t)$ will be unique
and $x$ as function of $b$ will be single-valued
with a jump, indicating a discontinuous phase transition, as shown
in figure \ref{apsv}.
In principle, the discontinuous transition could be
attained from the stationary probability distribution. 
Then, after taking the limit $t\to\infty$ followed by
$N\to\infty$, a single-valued with a jump could be obtained
\cite{qian2016b}. However, since we do not have an explicit
form of the probability distribution, we used 
the alternative method just explained.

\begin{figure}
\epsfig{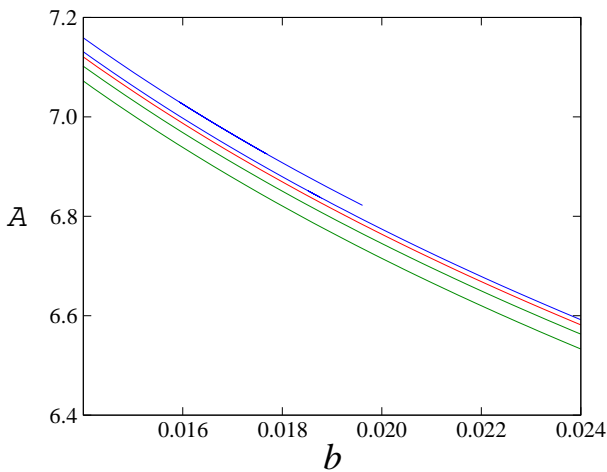}
\caption{The affinity ${\cal A}={\cal A}_1+{\cal A}_2$ as a function
of $b$. The parameters are the same as those of figure \ref{apsv}.}
\label{afin}
\end{figure}

\begin{figure}
\epsfig{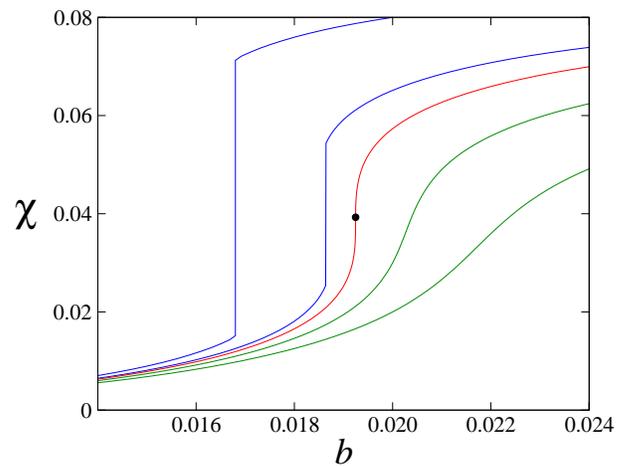}
\caption{The flux of reaction $\chi=\chi_1=\chi_2$ as a function of $b$.
The parameters are the same as those of figure \ref{apsv}.}
\label{fluxo}
\end{figure}

Figures \ref{apsv}, \ref{pe}, \ref{afin}, and \ref{fluxo}
show, respectively, the density $x$, the rate of
entropy production ${\sf P}$, the affinity ${\cal A}$, and 
the flux of reaction $\chi$ versus the activity $b$ for several
values of $a$, and for a set of the values of the rate constants.
For $a<a_c$, there is a discontinuous phase transition, indicated by
a jump in $x$. The rate of entropy production ${\sf P}$ and the
flux of reaction $\chi$ also display a jump
as shown in figure \ref{pe} and \ref{fluxo}.
Notice that, the activity ${\cal A}$ is continuous, in agreement
with the fact that it is a thermodynamic field.
At $a=a_c$, the jump in $x$ shrinks to zero
inducing the appearance of a critical point. At this point, the
rate of entropy production and the flux of reaction also become continuous.

\section{Conclusion}

We have analyzed reactive systems consisting of several chemical
reactions by the use of the stochastic thermodynamics. This approach
is based on a stochastic description of the time evolution of the
system, here described by a master equation,
and founded on two assumptions concerning the entropy.
The first being the definition of entropy according to Gibbs
and the other is the definition of entropy production based
on Schnakenberg expression, which is related to the ratio
of the probabilities of the forward and reverse trajectories 
in the space of microscopic states. We have shown that this approach
is fully connected to the energetics, being consistent with
thermodynamics. 

The stochastic trajectory occurs within a space constituted by
the microscopic states, which we choose to be the chemical
state of each particle. 
Under some circumstances, it is possible
to reduce to a description in terms of the number of particles
of each chemical species. In this case, the master equation
is reduced to the chemical master equation.  
By assuming that the equilibrium is
attained when the system is closed to particles, being
in contact with a heat reservoir only, we have obtained
relations that are obeyed by the transition rates of
each reaction. These relations partially defines the rates
and are used when the system is in contact with
particle reservoirs.

The reactive system is studied by placing it in contact 
with particle reservoirs, in addition to be in contact with
the heat reservoir. When the equilibrium condition given
by equation (\ref{22}) is not obeyed, the system will reach 
a nonequilibrium stationary state. In this case, there
will be fluxes of several types including fluxes of
particles and a flux of entropy which equals the entropy production.
This last quantity is written as a bilinear form in the
affinities and fluxes of particles, that is,
a sum of terms, each one 
being a product of the affinity of a reaction ${\cal A}_j$ 
and the flux of reaction $\chi_j$.
It should be remarked that this form was possible due to
the specific form of transition rates we have used  here.

We have focus mainly on the production of entropy and applied
our approach to the first and second Schl\"ogl models.
The second model displays a discontinuous phase transition
and a critical point. The density, the particle flux and the
production of entropy show a jump at the transition being
continuous at the critical point. We remark that the
affinities is continuous, which is consistent with the fact 
that it is a thermodynamic field, usually called intensive variable.

We have shown that the bilinear form of entropy, given by
equation (\ref{69a}), can also be written in the form (\ref{69b}).
Usually,
this is the expression used to determine the entropy production
within the chemical kinetic approach. Although both these
formulas give the same result for the entropy, they are
conceptually distinct due to the presence of the affinity,
which is a thermodynamic field, in the bilinear form (\ref{69a}).
This form is the one appropriate to get for instance the Onsager
coefficients.



\begin{thebibliography}{99}

\bibitem{avery1974} H. E. Avery, {\it Basic Reaction Kinetics and Mechanisms},
Macmillan, London, 1974.

\bibitem{yeremin1979} E. N. Yeremin, {\it The Foundations of Chemical Kinetics}
Mir, Moscow, 1979.

\bibitem{houston2001} P. L. Houston, {\it Chemical Kinetics and Reaction Dynamics},
McGraw-Hill, New Yor, 2001.


\bibitem{nicolis1977} G. Nicolis and I. Prigogine, {\it Self-Organization in 
Nonequilibrium Systems}, Wiley, New York, 1977.

\bibitem{kampen1981} N. G. van Kampen,
{\it Stochastic Processes in Physics and Chemistry},
North-Holland, Amsterdam, 1981.

\bibitem{tomebook2015} T. Tom\'e and M. J. de Oliveira,
{\it Stochastic Dynamics and Irreversibility},
Springer, Heidelberg, 2015.


\bibitem{mcquarrie1967} D. A. McQuarrie,
J. Appl. Prob. {\bf 4} 413 (1967).

\bibitem{kurtz1972} T. G. Kurtz, 
J. Chem. Phys. {\bf 57}, 2976 (1972).

\bibitem{gillespie1992} D. T. Gillespie, Physica A {\bf 188}, 404 (1992).

\bibitem{gillespie2000} D. T. Gillespie,
J. Chem. Phys. {\bf 113}, 297 (2000).


\bibitem{jiuli1984} L. Jiu-Li, C. Van den Broeck and G. Nicolis,
Z. Phys. B {\bf 56},165 (1984).

\bibitem{mou1986} C. Y. Mou, J.-L. Luo and G. Nicolis,
J. Chem. Phys. {\bf 84}, 7011 (1986).

\bibitem{horowitz2015} J. M. Horowitz,
J. Chem. Phys. {\bf 143}, 044111 (2015).

\bibitem{ge2016} H. Ge and H. Qian,
Chem. Phys. {\bf 472}, 241 (2016).

\bibitem{qian2016a} H. Qian, S. Kjelstrup, A. B. Kolomelsky and D. Bedeaux,
J. Phys.: Condens. Matter {\bf 28}, 153004 (2016)

\bibitem{tome2006} T. Tom\'e, Braz. J. Phys. {\bf 36}, 1285 (2006).

\bibitem{tome2010} T. Tom\'e and M. J. de Oliveira,
Phys. Rev. E {\bf 82}, 021120 (2010).

\bibitem{tome2012} T. Tom\'e and M. J. de Oliveira,
Phys. Rev. Lett. {\bf 108}, 020601 (2012). 


\bibitem{tome2015} T. Tom\'e and M. J. de Oliveira,
Phys. Rev. E {\bf 91}, 042140 (2015).

\bibitem{zia2006} R. K. P. Zia and B. Schmittmann,
J. Phys. A: Math. Gen. {\bf 39}, L407 (2006).

\bibitem{zia2007} R. K. P. Zia and B. Schmittmann,
J. Stat. Mech. P07012 (2007).

\bibitem{schmiedl2007} T. Schmiedl and U. Seifert,
J. Chem. Phys. {\bf 126}, 044101 (2007).

\bibitem{seifert2008} U. Seifert, Eur.Phys. J. B {\bf 64}, 423 (2008).

\bibitem{esposito2009} M. Esposito, K. Lindenberg, and
C. Van den Broeck, Phys. Rev. Lett. {\bf 102}, 130602 (2009).

\bibitem{broeck2010} C. Van de Broeck and M. Esposito,
Phys. Rev. E {\bf 82}, 011144 (2010).

\bibitem{esposito2012} M. Esposito, 
Phys. Rev. E {\bf 85}, 041125 (2012).

\bibitem{seifert2012} U. Seifert, Rep. Prog. Phys. {\bf 75},
126001 (2012).

\bibitem{zhang2012} X.-J. Zhang, H. Qian and M. Qian,
Phys. Rep. {\bf 510}, 1 (2012).

\bibitem{ge2012} H. Ge, M. Qian and H. Qian,
Phys. Rep. {\bf 510}, 87 (2012).

\bibitem{luposchainsky2013} D. Luposchainsky and H. Hinrichsen,
J. Stat. Phys. {\bf 153}, 828 (2013).

\bibitem{wu2014} W. Wu and J. Wang,
J. Chem. Phys. {\bf 141}, 105104 (2014).


\bibitem{prigogine1947} I. Prigogine, {\it Etude Thermodynamique des
Ph\'enom\`enes Irr\'eversibles}, Desoer, Li\`ege, 1947.

\bibitem{prigogine1950} I. Prigogine et R. Defay, {\it Thermodynamique
Chimique}, Desoer, Li\`ege, 1950.

\bibitem{prigogine1955} I. Prigogine, {\it Introduction to Thermodynamics
of Irreversible Processes}, Thomas, Springfield, 1955.

\bibitem{groot1962} S. R. de Groot and P. Mazur,
{\it Non-Equilibrium Thermodynamics},
North-Holland, Amsterdam, 1962.

\bibitem{glansdorff1971} P. Glansdorff and I. Prigogine, {\it Thermodynamics
of Structure, Stability and Fluctuations}, Wiley, New York, 1971.


\bibitem{donder1927} Th. De Donder, {\it L'Affinit\'e},
Lamertin, Bruxelles, 1927.

\bibitem{donder1922} Th. De Donder,
Bulletin de la Classe des Sciences, Acad\'emie Royale de Belgique
{\bf 8}, 197-205 (1922).

\bibitem{donder1925} Th. De Donder,
Comptes Rendue Hebdomadaires des S\'eances de l'Acad\'emie des Sciences
{\bf 180}, 1334-1337 (1925).


\bibitem{clausius1865} R. Clausius,
Annalen der Physik und Chemie 125, 353–400 (1865).


\bibitem{schnakenberg1976} J. Schnakenberg,
Rev. Mod. Phys. {\bf 48}, 571 (1976).


\bibitem{decker2016} Y. De Decker, J.-F. Derivaux, and G. Nicolis,
Phys. Rev. E {\bf 93}, 042127 (2016).


\bibitem{arrhenius1889} S. A. Arrhenius,
Z. Phys. Chem. {\bf 4}, 96 (1889); {\bf 4}, 226 (1889).

\bibitem{moore1965} W. J. Moore, {\it Physical Chemistry},
Longman, London, 1965.

\bibitem{callen1960} H. B. Callen, {\it Thermodynamics},
Wiley, New York, 1960.

\bibitem{reichl1980} L. E. Reichl, {\it A Modern Course in 
Statistica Mechanics}, Universiy of Texas Press, Austin, 1980.

\bibitem{kondepudi1998} D. Kondepudi and I. Prigogine,
{\it Modern Thermodynamics}, Wiley, New York, 1998.

\bibitem{oliveira2013} M. J. de Oliveira, {\it Equilibrium Thermodynamics},
Springer, Heidelberg, 2013.


\bibitem{lewis1923} G. N. Lewis and M. Randall, {\it Thermodynamics
and the Free Energy of Chemical Substances},
McGraw-Hill, New York, 1923.


\bibitem{onsager1931} L. Onsager,
Phys. Rev. {\bf 37}, 405 (1931); {\bf 38}, 2265 (1931).


\bibitem{schlogl1972} F. Schl\"ogl, 
Z. Phys., {\bf 253} 147 (1972).

\bibitem{gaspar2004} P. Gaspard,
J. Chem. Phys, {\bf 120} 8898 (2004).

\bibitem{vellela2009} M. Vellela, H. Qian,
J. R. Soc. Interface, {\bf 6}, 925 (2009)

\bibitem{endres2015} R. G. Endres,
PLoS One {\bf 10}, 0121681 (2015)

\bibitem{endres2017} R. G. Endres,
Scientific Reports {\bf 7}, 14437 (2017).

\bibitem{qian2016b} H. Qian, P. Ao, Y. Tu and J. Wang,
Chem. Phys. Lett. {\bf 665}, 153 (2016).

\end{thebibliography}
\end{document}